 \pdfoutput=1
\documentclass[floatfix]{emulateapj}
 \usepackage{amsmath}
\usepackage[boxed]{algorithm2e}
\usepackage{algorithmic}
\usepackage{epsfig,subfigure,color,amsmath,graphicx,multirow,subfigmat,ctable,bm}

\newcommand{\beq}{\begin{equation}}
\newcommand{\eeq}{\end{equation}}
\newcommand{\beqs}{\begin{eqnarray}}
\newcommand{\eeqs}{\end{eqnarray}}

\newcommand{\norm}[1]{\left\lVert#1\right\rVert}
\input{epsf} 
\newcommand{\bfi}{\begin{figure} \epsfxsize=8cm \epsffile}
\newcommand{\efi}{\end{figure}}

\begin{document}
\title{Non-negative matrix factorization for self-calibration of photometric redshift scatter in weak lensing surveys}
\author{Le Zhang$^{1}$, Yu Yu$^{1,2}$, Pengjie Zhang$^{1,3,4}$\\
{~}\\
$^{1}$ Department of Astronomy, Shanghai Jiao Tong University, Shanghai, 200240\\
$^{2}$Key laboratory for research in galaxies and cosmology, Shanghai Astronomical Observatory, Chinese Academy of Sciences, 80 Nandan Road, Shanghai, China, 200030\\
$^{3}$IFSA Collaborative Innovation Center, Shanghai Jiao Tong
University, Shanghai 200240, China\\
$^{4}$ Tsung-Dao Lee Institute, Shanghai 200240, China
}
\thanks{Email: lezhang@sjtu.edu.cn}
\begin{abstract}
Photo-$z$ error is  one of the major sources of systematics degrading the accuracy of weak lensing cosmological inferences. 
\cite{2010MNRAS.405..359Z} proposed a self-calibration method  combining galaxy-galaxy correlations and galaxy-shear correlations  between different photo-$z$ bins. Fisher matrix analysis shows that  it can determine the rate of photo-$z$ outliers at a level of $0.01$-$1\%$ merely using photometric data and do not rely on any prior knowledge. In this paper, we develop a new algorithm to implement this method by solving a constrained nonlinear optimization problem arising in the self-calibration process. Based on the techniques of fixed-point iteration and non-negative matrix factorization, the proposed algorithm can efficiently and robustly reconstruct the scattering probabilities between the true-$z$ and photo-$z$ bins. The algorithm has been tested extensively by applying it to mock data from simulated stage IV weak lensing projects. We find that the algorithm provides a successful recovery of the scatter rates at the level of $0.01$-$1\%$, and the true mean redshifts of photo-$z$ bins at the level of $0.001$, which may satisfy the requirements in future lensing surveys.

\end{abstract}


\maketitle

\section{Introduction}
Weak gravitational lensing is one of the most powerful cosmological probes to study the distribution of dark matter, the dynamics of dark energy and the formation of large scale structures in the universe, as well as the nature of gravity at cosmological scales (e.g. ~\citet{2003ARA&A..41..645R,Albrecht:2006um,2008PhRvD..78f3503J,2013PhR...530...87W,2015APh....63...23H}). By statistically measuring the cosmic shear signals, existing surveys, including CFHTLS\footnote{http://www.cfht.hawaii.edu/Science/CFHLS/}~\citep{Kilbinger:2012qz,Heymans:2013fya,2014MNRAS.441.2725F} and SDSS~\citep{2014MNRAS.440.1322H,2013MNRAS.432.1544M}, have already tightly constrained a combination of the amplitude of density fluctuation $\sigma_8$ and the matter density parameter $\Omega_m$. Ongoing weak lensing surveys such as DES\footnote{https://www.darkenergysurvey.org/} and HSC\footnote{http://www.naoj.org/Projects/HSC/index.html}, along with the planned stage IV surveys such as  Euclid\footnote{http://sci.esa.int/euclid/} and LSST\footnote{http://www.lsst.org/lsst/}, will further increase the survey depth to $z_S>1$, and increase the available number of source galaxies by orders of magnitude. 

Despite the great capability of these surveys to achieve high precision weak lensing measurements, weak lensing cosmology suffers from various sources of systematic errors such as  shape measurement errors ~\citep{Heymans:2005rv,Massey:2006ha,2015MNRAS.450.2963M} and   intrinsic alignments of source  galaxies~\citep{2004MNRAS.353..529H,2006MNRAS.367..611M,2007MNRAS.381.1197H,2009ApJ...694..214O,2009ApJ...694L..83O,2015PhR...558....1T}. Another major systematic error is the photometric redshift (photo-$z$) errors~\citep{2006ApJ...636...21M,2010MNRAS.401.1399B}. Even though the telescope technology developments have made rapid progress, spectroscopic redshift surveys of a large number of galaxies are still very time consuming especially for high redshift large scale galaxy surveys. All of the above-mentioned large galaxy surveys will thus have photometric rather than spectroscopic redshift identifications. This results in  non-negligible  photo-$z$ errors. It is found that both the bias and scatter need to be controlled at the level of $10^{-3}$ to avoid a considerable degradation in cosmological parameter accuracies~\citep{2006ApJ...636...21M,2006MNRAS.366..101H,2008MNRAS.389..173K}. Therefore, it is crucial to precisely determine the true redshift distribution of galaxies in a photometric weak lensing survey.

Besides efforts of directly improving the photo-$z$ estimation algorithms, attempts have been made to infer the true redshift distribution of source galaxies by the large scale structure (LSS) statistics.  For example, the cross-correlation techniques that measure redshift errors by cross correlating the photo-$z$ samples with spectroscopic ones in overlapping survey areas have been proposed recently \citep{2008ApJ...684...88N,2010ApJ...721..456M,2010ApJ...724.1305S,2012ApJ...745..180M,2013arXiv1303.4722M,2013MNRAS.431.3307S,2013MNRAS.433.2857M}. It has been applied in various surveys (e.g. \citet{2012ApJ...753...23M,2016MNRAS.457.3912R,2016MNRAS.460..163R,2016MNRAS.462.1683S,2016arXiv161107578J,2016MNRAS.463.3737C}). This cross-calibration technique is powerful. However, it suffers from two potentially significant systematic errors. For spectroscopic galaxies in the redshift bin $z_S$, its cross correlation with the photometric galaxies of a given photo-$z$ bin is $\propto b_S(z_S)b_P(z_S)r_{SP}(z_S) P_{P\rightarrow S}$. Here, $b_S$ is the bias of spectroscopic galaxies.  $b_P(z_S)$ is the bias of photometric galaxies whose true redshifts fall in the redshift bin $z_S$,  $P_{P\rightarrow S}$ is the fraction of these photometric galaxies, and $r_{SP}$ is the cross-correlation coefficient between these galaxies and the spectroscopic galaxies at $z_S$. $P_{P\rightarrow S}$ quantifies the photo-$z$ error, but it is degenerate with $b_P(z_S)$ and $r_{SP}(z_S)$. The degeneracy with $b_P$ has been investigated~\citep{2014ApJ...780..185D,2015MNRAS.447.3500R}. It can result in significant systematic error in the determined photo-$z$ or even complete failure~\citep{2014ApJ...780..185D}. The second degeneracy is seldom addressed in the literature. But even a small stochasticity ($1-r_{SP}\sim \mathcal{O}(0.01)$) can result in $\mathcal{O}(1\%)$ systematic error in $P_{P\rightarrow S}$. Therefore it severely limits the application of cross-calibration in the nonlinear regime where we expect significant stochasticity.


Meanwhile,  the self-calibration techniques, which does not rely on external spectroscopic data, have been proposed and applied in real data \citep{2006ApJ...651...14S,2009A&A...493.1197E,2010MNRAS.405..359Z,2010MNRAS.408.1168B,2010ApJ...725..794Q}. Under the Limber approximation, the galaxy-galaxy angular correlation between two different redshift bins should vanish if there are no photo-$z$ errors. Therefore,  the galaxy-galaxy correlation between redshift bins can be used to infer the photo-$z$ outlier rate. The galaxy-galaxy correlation can be measured to high precision by future weak lensing surveys, leading to robust constraint on photo-$z$ error. However, an intrinsic degeneracy between up and down scatters would severely limit the accuracy of the determination of outlier rates, since a non-zero galaxy-galaxy correlation between redshift bins can be induced by both up and down scatters and galaxy-galaxy measurement alone can not break this degeneracy~\citep{ 2010MNRAS.405..359Z,2010MNRAS.408.1168B}. 


\citet{2010MNRAS.405..359Z} pointed out that the galaxy-shear angular correlation available in the same weak lensing survey naturally breaks this degeneracy, leading to significant or even orders of magnitude improvement in the calibration accuracy. This establishes the galaxy-shear correlation as an indispensable part of the self-calibration technique. The self-calibration technique does not rely on any assumptions on cosmological priors, galaxy bias and parameterization of the photo-$z$ probability distribution function. Through the measurements of galaxy-galaxy correlation and galaxy-shear correlation between different photo-$z$ bins, it can accurately reconstruct  the full relation between the photometric and spectroscopic redshift and, for a given photo-$z$ bin, derive the fraction of galaxies which in fact come from a distinct true redshift bin.  For a stage IV projects like LSST, it can detect outliers with rate as low as $0.01\%$-$1\%$.  

The theoretical reconstruction accuracy of such method is obtained from the Fisher matrix analysis \citep{2010MNRAS.405..359Z}. In practice, however, implementing such calibration process requires solving the set of nonlinear constrained matrix equations, which is a computationally hard problem. So far, implementations in data analysis such as the pairwise analysis \citep{2009A&A...493.1197E,2010MNRAS.408.1168B} are based on significant simplifications of these equations and therefore can not meet the requirement of stage IV projects. 

The exact and accurate implementation of the self-calibration is thus the focus of this paper. Unfortunately, none of the widely used algorithms proposed in the literature is capable of finding a reliable solution that are close enough to the simulation truth,  e.g., the Powell's method, the quasi-Newton method and expectation-maximization method. 
 Even for Monte Carlo-based fitting methods, the estimation of a large number of parameters, typically in the order of 100, to the desired accuracy is still prohibitive computationally. The reason is not only the non-convex nonlinear optimization problem with numerous unknown parameters need to be solved, but also multiple constraints have to be imposed for those unknowns.
   
In this paper, we report on the development of a novel algorithm that can be implemented in the self-calibration process. The algorithm is based on the techniques of fixed-point iteration and non-negative matrix factorization. In Sect.~\ref{sect:Form}, we briefly review the self-calibration theory and give a detailed description about the proposed algorithm. Sect.~\ref{sect:test} presents the application of this algorithm to mock photometric data and shows the main results. Finally, Sect.~\ref{sect:con} provides discussion and concluding remarks.

The notations and conventions used in this paper are as follows.
The superscript ``P'' denotes a property in photo-$z$ bins, and the superscript ``R'' denotes a corresponding property in true-$z$ bins. The capital ``G'' denotes gravitational lensing, to be more specific, the lensing convergence converted from the more direct observable cosmic shear, and the little ``g'' denotes galaxy number density (or over-density). Furthermore, upper case letters are used to denote matrices. let $A$ be a matrix. Then the $(i,j)$-th entry of a matrix A is referred to by either $A_{ij}$ or $a_{ij}$, and its transpose by $A^T$. 

\section{Problem Formalization}\label{sect:Form}
As proposed by~\cite{2010MNRAS.405..359Z}, the rate of photo-$z$ outliers can be determined by using galaxy-galaxy and lensing-galaxy measurements in photometric data from the original lensing survey. Let us start with the measurement equations in terms of photo-$z$ scatters, 
\beqs
C^{gg,P}_{ij}(\ell) &=& \sum_k P_{ki}P_{kj} C^{gg,R}_{kk}(\ell)+ \delta N^{gg,P}_{ij}(\ell)\,,\\
C^{Gg,P}_{ij}(\ell) &=& \sum_{k\geq m} P_{ki}P_{mj} C^{Gg,R}_{km}(\ell)+\delta N^{Gg,P}_{ij}(\ell)\,,
\eeqs
where $P_{ij}$ represents the scattering probability between the $i$-th true-$z$ bin and the $j$-th photo-$z$ bin. To be exact, $P_{ij}=N_{i\rightarrow j}/N_j^P$, where $N_j^P$ is the total number of galaxies in the $j$-th photo-$z$ bin and $N_{i\rightarrow j}$ of them come from the $i$-th true-$z$ bin.  $P_{ij}$ satisfies two conditions,  $P_{ij}\geq 0$ and $\sum_i P_{ij} =1$. Notice that the first subscript of $P_{ij}$ denotes the label of true-$z$ bin and the second subscript denotes the label of photo-$z$ bin. Therefore it is in general asymmetric ($P_{ij} \neq P_{ji}$). $C^{gg,P}_{ij}$ stands for the measured galaxy power spectrum between the $i$-th and $j$-th photo-$z$ bins, and $C^{Gg,P}_{ij}$ denotes the measured cross correlation power spectrum between the lensing convergence in the $i$-th photo-$z$ bin and the galaxy number density in the $j$-th photo-$z$ bin. As defined above, $C^{gg,R}_{ij}$ and $C^{Gg,R}_{ij}$ are the corresponding power spectra in the true redshift bins. For a given $\ell$, $C^{gg,R}$ is expected to be $C^{gg,R}_{k\neq m} =0$, as the galaxy cross correlation between non-overlapping redshift bins would vanish under the Limber approximation. Furthermore, using the notation where a larger index corresponds to a higher redshift, we also expect $C^{Gg,R}_{k\geq m} \neq 0$ and $C^{Gg,R}_{k< m} = 0$. This is because that, in the absence of lensing magnification bias, only source galaxies behind a lens can be lensed, resulting in non-zero lensing-galaxy correlations. Therefore, we are only summing over the diagonal ($k=m$) and lower triangular ($k\geq m$) components in the above measurement equations.

In realistic cases, the measured power spectra are certainly contaminated by shot noise, so that we introduce the shot noise terms, $\delta N^{gg,P}$ and $\delta N^{Gg,P}$, which represent the fluctuations of the associated shot noise in measurements. Note that here we have implicitly subtracted the ensemble average of shot noise out of the observations as it can not bias the estimate of outliers. That is the reason why only the fluctuations $\delta N^{gg,P}$ and $\delta N^{Gg,P}$ are taken into account in the measurement equations.

For a given $\ell$, if we split galaxies into $n$ photo-$z$ bins, the same equations but in matrix notation read
\beqs
C^{gg,P}_\ell &=& P^T  C^{gg,R}_\ell P +\delta N^{gg,P}_\ell \,,\label{eq:Cgg}\\
C^{Gg,P}_\ell &=& P^T C^{Gg,R}_\ell P + \delta N^{Gg,P}_\ell\,,\label{eq:CGg}
\eeqs
where $P\in \mathbb{R}^{n\times n}$ is a non-symmetric matrix, subject to the column-sum-to-one constraint, i.e., $\sum_i P_{ij}=1$, for all $j$. The matrix $P$ is also a so-called non-negative matrix, that is, $P_{ij}\geq 0$ for all $(i,j)$.  $C^{gg,R}_\ell\in \mathbb{R}^{n\times n}$ is a non-negative diagonal matrix, and $C^{Gg,R}_\ell\in \mathbb{R}^{n\times n}$ is a lower triangular matrix.

For a given data set, $C^D=\{C^{gg,P},C^{Gg,P}\}$, the parameters, $\theta = \{P, C^{gg,R}_\ell, C^{Gg,R}_\ell\}$, can be estimated by the posterior parameter distribution $P(\theta|C^{D})$. Using Bayes' rule, $P(\theta|C^{D})$ can be expressed by $P(\theta|C^{D}) \propto {\cal L}(C^{D}|\theta) P(\theta)$, where the likelihood function of the data is ${\cal L}(C^{D}|\theta)$ and the parameter prior is $P(\theta)$. Assuming a flat prior on the parameters, we will see that the best-fitted parameters can be derived by maximizing the likelihood function. 
 
As the fluctuations are induced by shot noise which is assumed to be uncorrelated at different $z^P$ bins and $\ell$ bins, different power spectrum measurements thus are assumed to be uncorrelated and can be well approximated by a Gaussian distribution if there are sufficiently large number of independent $\ell$ modes in each multipole $\ell$ bin. With these assumptions, the likelihood function then becomes a simple multivariate Gaussian distribution with the covariance matrix $\Sigma_N$, i.e., $P(C^{D}|\theta)= {\cal N}(C^{D}-\widehat{C}(\theta); \Sigma_N)$. Each element in $\Sigma_N$ is evaluated by $\left<\delta N^{\alpha,P}_{i,j}(\ell) \delta N^{\beta,P}_{i',j'}(\ell') \right> = \left[\sigma^\alpha_{ij}\right]^2\delta_{\alpha,\beta}\delta_{i,i'}\delta_{j,j'}\delta_{\ell,\ell'}$, with $\alpha, \beta = gg, Gg$, and     
\beqs\label{eq:noise1}
\big(\sigma^{gg,P}_{ij}\big)^2 &=& \frac{1}{(2\ell+1)\Delta \ell f_{\rm sky} } \frac{1}{\bar{n}_i\bar{n}_j}(1+\delta_{ij}) \,,\\
\big(\sigma^{Gg,P}_{ij}\big)^2 &=& \frac{1}{(2\ell+1)\Delta \ell f_{\rm sky} } \frac{\gamma^2_{\rm rms}}{\bar{n}_i\bar{n}_j}\label{eq:noise2} \,,
\eeqs
for data in bins of width $\Delta \ell$, where $f_{\rm sky}$ is the fraction of the sky observed, $\bar{n}_i$ is the mean galaxy surface density in the $i$-th redshift bin per steradian and $\gamma_{\rm rms}$ denotes the rms dispersion in the shear measurement induced by the galaxy intrinsic ellipticities.

It is straightforward to show that the logarithm of the likelihood function can be finally simplified to 

\beqs\label{likelihood}
\ln {\cal L} \propto &&-\sum_{\ell} \bigg[\sum_{i\leq j} \left(\sigma^{gg}_{ij}(\ell)\right)^{-2} \left(C^{gg,P}_{ij}(\ell)- \widehat{C}^{gg,P}_{ij}(\ell) \right)^2  \nonumber \\
&&+\sum_{ij} \left(\sigma^{Gg}_{ij}(\ell)\right)^{-2} \left(C^{Gg,P}_{ij}(\ell)- \widehat{C}^{Gg,P}_{ij}(\ell)\right)^2 \bigg]\,,\quad
\eeqs 
where $\widehat{C}^{gg,P}_{ij}(\ell) \equiv P^T C^{gg,R}_\ell P  $ and $\widehat{C}^{Gg,P}_{ij}(\ell)\equiv  P^T C^{Gg,R}_\ell P $ are the estimated quantities in terms of ($P, C^{gg,R}_\ell, C^{Gg,R}_\ell$).

Note that, in an ideal case that the noise is completely neglected, the model-independent self-calibration technique requires solving the set of nonlinear matrix equations given by Eqs.~\ref{eq:Cgg} \& \ref{eq:CGg} for all $P$, $C^{gg,R}_\ell$ and $C^{Gg,R}_\ell$ simultaneously with the required constraints. In the next section, we will first present a new algorithm for solving such non-linear problem in the idealized noise-free case, and then will come back to the realistic situation in Sect.~\ref{sect:opt}.

\subsection{Fixed-point-based  algorithm}\label{sect:fp}
In the absence of shot noise, the data depend deterministically on the matrices $P$, $C^{gg,R}_\ell$ and $C^{Gg,R}_\ell$. As the number of measurements in many multipole $\ell$ bins is always much larger than that of the unknown parameters, in principle, the unique exact solution exists for Eqs.~\ref{eq:Cgg} \& \ref{eq:CGg}. However, due to quadratic dependence on $P$ and linear on $C^{gg,R}_\ell$ and $C^{Gg,R}_\ell$, the analytical solution remains unknown for such complex system. We therefore resort to a numerical method.

The observation that, the problem can be split into two subproblems, suggests a method for solving the two subproblems independently. According to the form of Eq.~\ref{eq:Cgg} where $C^{gg,R}_\ell$ is a diagonal matrix, one can find a matrix, which has rows of ``true'' $P$ but in permuted sequence, also satisfies Eq.~\ref{eq:Cgg}. On the other hand , Eq.~\ref{eq:CGg} alone can uniquely determine the order of the rows since only the ``true'' $P$ will make the reconstructed $\widehat{C}_\ell^{Gg,R}$ to be lower triangular matrices, which implies all of the remaining elements, i.e., $[C^{Gg,R}_\ell]_{ij}$ with $i<j$, are exactly zeros. The reconstruction error thus can be quantitatively measured by 
\beq
{\epsilon}^{Gg} \equiv  \sum_\ell \sum_{i<j} \sqrt{([\widehat{C}^{Gg,R}_\ell]_{ij})^2} \label{eq:ecGg}\,
\eeq
where $[\widehat{C}^{Gg,R}_\ell]_{ij} =\sum_{km} [P^{-1}]_{ki} [C^{Gg,P}_\ell]_{km} [P^{-1}]_{mj}$

For these concerns, we develop a novel fixed-point-based algorithm, which is capable of solving such high dimensional non-linear problem iteratively. We first solve for the relatively simpler problem of Eq.~\ref{eq:Cgg}, and then the true $P$ can be obtained by utilizing Eq.~\ref{eq:CGg} to break down the permutation degeneracy. The procedure of the algorithm is summarized in Algorithm~\ref{algo1}.

\begin{algorithm}[htpb]
  \caption{The fixed-point-based algorithm for solving Eqs. \ref{eq:Cgg} \& \ref{eq:CGg} in the absence of shot noise.}
  \label{algo1} 
\begin{algorithmic} 
\STATE \COMMENT {\hrulefill\ {\bf Step 1:} solving for Eq.~\ref{eq:Cgg}\hrulefill\  }
\REQUIRE{Non-negative data matrices $C^{gg,P}_\ell$ for all $\ell$.} 
\STATE {\bf initialize}: \\
(a) Assign $P$ to a random positive matrix with \\
~~~~~$\sum_i P_{ij}=1$, for all $i$\\
(b) Set $C^{gg,R}_\ell = Abs\{ {P^{T}}^{-1} C^{gg,P}_\ell P^{-1}\}$, for all $\ell$ \\
\REPEAT 
 
\FOR {$\ell=1$ to $n_\ell$} 
\STATE $C^{gg,R}_\ell = {P^T}^{-1}C^{gg,P}_\ell P^{-1}$
\STATE $Q_\ell = C^{gg,P}_\ell P^{-1}$
\STATE $P^T =Abs\{(\sum_{\ell} Q_\ell) (\sum_{\ell}C^{gg,R}_\ell)^{-1} \}  $
 \STATE $v_j= \sum_k [P^T]_{jk}$, for all $j$ 
\STATE $[P^T]_{ij} = \frac{[P^T]_{ij}+\epsilon}{v_{j}+\epsilon}$, for all $(i,j)$ \% normalization 
\ENDFOR

\STATE $C^{gg,R}_\ell = Abs\{Diag\{C^{gg,R}_\ell\}\}$, for all $\ell$

\UNTIL {a convergence criterion is satisfied}
\ENSURE{$P$} 

\STATE \COMMENT {\hrulefill\ {\bf Step 2:} determine the row order of $P$\ \hrulefill}
\REQUIRE{Data matrices $C^{Gg,P}_\ell$ for all $\ell$, and $P$ from Step 1}
\REPEAT 
\STATE 1. Randomly swap two rows of $P$ 
\STATE 2. Compute $\epsilon^{Gg}$ with Eq.~\ref{eq:ecGg}
\UNTIL {$\epsilon^{Gg}\leq \epsilon$  (a typical value of $\epsilon$ is $10^{-10}$)} 
\ENSURE{$P$} 

\end{algorithmic}
\end{algorithm}

Note that, there are two important tricks implemented into Algorithm~\ref{algo1}. One trick is that $P$ is successively updated by averaged $Q_\ell$ and $C^{gg,R}_\ell$ overl all $\ell$, similar to the successive over relaxation method. The averaging scheme allows a remarkable convergence rate, and can make a solution significantly stabilized to desired precision after some iterations, which would otherwise diverge. 

The other one is to introduce the vectors $v_j$, which is based on fixed-point method and is essential for this algorithm. This method has two-fold advantages as follows. 

Firstly, in order to deal with the constraint in $P$, the commonly used normalization step can be imposed during iterations to enforce the unitary sum for each column of $P$. However, we find such normalization step may cause unstable and slow convergence. Instead, we introduce vectors $v_j$ for rescaling each element of columns in $P$ with different values, which can be regarded as slight perturbations in $P$. It is straightforward to see that all of the elements in the rescaling vectors are unity when $P$ satisfies Eq.~\ref{eq:Cgg} with the column-sum-to-one constraint. For this reason, the perfect recovery is necessarily a fixed point of the update rules. 

Secondly, as we know, the solution obtained by an iterative process may prematurely get stuck in an undesirable local maximum before reaching the global maximum (i.e., the true solution), especially for a constrained optimization problem with a large number of parameters. Since the sum of each column of $P$ is not exactly unity before $P$ converges to the true value, the vectors $v_j$ can slightly relax the column-sum-to-one constraint to overcome local maxima.
    
Simulation results are given in Sect.~\ref{sect:test} and will show that the fixed-point-based algorithm converges fast, and can efficiently solve the set of nonlinear matrix equations in an iterative way until the desired accuracy is attained.

\subsection{Optimization algorithm}\label{sect:opt}
In reality, observations are corrupted by measurement noise, and hence the fixed-point-based algorithm may not be desirable for a direct application to noisy data. 

The standard approach for parameter estimate under a given noisy data model is by maximizing the likelihood function. Thanks to the fact that a maximum-likelihood estimate of the parameters is equivalent to a least-squares estimate as long as the errors for all data points are independent and belong to a same Gaussian distribution. The measurement noise in weak lensing surveys is usually assumed to be a shot noise well approximated by a Gaussian distribution. If we appropriately choose the bin-widths of multipole $\ell$ and redshift $z$, the noise levels $\sigma^{gg}_{ij}(\ell)$ in the measured galaxy-galaxy power spectrum would be approximately identical over all $\ell$ bins. From Eq.~\ref{eq:noise1}, we can find that the estimated noise levels have a factor of 2 difference between the diagonal ($i=j$) and off-diagonal ($i\neq j$) components, which would make the least-squares estimate slightly deviate from the maximum-likelihood estimate. However, such deviation is expected to be negligible as the noise levels are very small quantities compared to $C^{gg,P}_{\ell}$, if one choose bin-widths of $\Delta \ell$ and $\Delta z$ sufficiently large. Therefore, the problem in Eq.~\ref{eq:Cgg} can be straightforwardly reformulated as the following optimization problem:
\beqs
\text{min}&& {\cal J}(P;C^{gg,R}_{\ell=1,\ldots,n_\ell}) \equiv \frac{1}{2}\sum_\ell \norm{C^{gg,P}_\ell -P^T  C^{gg,R}_\ell P}_F^2,\quad\quad\label{eq:cggmin}\\
s.t.~~ &&\left\{\begin{aligned}
&P,\ C^{gg,R}_\ell \geq 0\,, \text{\,for all~}\, \ell  \nonumber \\
&\sum_i P_{ij}=1\,, \text{\,for all~}\, j \nonumber \\
&C^{gg,P}_\ell = \mathrm{Diag}\{C^{gg,P}_\ell\},\, \text{\,for all~}\,  \nonumber \ell \nonumber 
\end{aligned}
\right.
\eeqs
where $\norm{.}$ is the Frobenius norm. We aim at minimizing the objective function ${\cal J}$ which measures a accumulated decomposition error across all data matrices. Since we seek to represent each non-negative data matrix as the product of non-negative matrices, Non-negative Matrix Factorization (NMF) is well-suited  to solve this constrained optimization problem. 
\subsubsection{Non-negative Matrix Factorization}
In this section, we will first briefly review the background of the NMF method, and then describe the proposed approach based on NMF for solving the optimization problem in Eq.~\ref{eq:cggmin}. 

Given an arbitrary $m\times n$ non-negative matrix $V$, NMF finds non-negative matrices $W$ and $H$ such that
\beq
\text{min}   \norm{V-WH}_F^2\quad s.t.~ W,~H \geq 0\,,
\eeq
where $W\in \mathbb{R}^{m\times r}$, $H\in \mathbb{R}^{r\times n}$. \cite{Lee01algorithmsfor} has found the following ``multiplicative update rules'' to minimize this conventional least squares error and also the error is not nonincreasing during iterations under this update rules:
\beq
W_{ij} \leftarrow W_{ij}\frac{(VH^T)_{ij}}{(WHH^T)_{ij}} \quad H_{ij} \leftarrow H_{ij}\frac{(W^TV)_{ij}}{(W^TWH)_{ij}}\,. 
\eeq

For constructing the multiplicative update rules adapted to the optimization problem in Eq.~\ref{eq:cggmin}, the most difficult part is that the objective is quartic with respect to $P$. To circumvent this situation, the natural way for updating $P$ is to translate this tri-factor NMF to the standard bi-factor NMF. For example, one can use a simple so-called ``splitting'' technique. That is, the two appearances of $P$ in Eq.~\ref{eq:cggmin} are represented by two different matrices, say $P_L$ and $P_R$, which are optimized independently. The optimization problem now amounts to minimizing 
\beq\label{eq:minimize}
\frac{1}{2}\sum_\ell \norm{C^{gg,P}_\ell -P_L^T  C^{gg,R}_\ell P_R}_F^2\,.
\eeq
Therefore, one can update $P_L$ and $P_R$ alternatively and iteratively. After convergence of a series of iterations, it is hoped that those two matrices obtained happen to be equal. However, we find that this is not always the case in practice for this specific problem. 

Since we know $P_L$ and $P_R$ should be equal when the correct solution is obtained, enforcing $P_L =R_R$ after the update of either $P_L$ or $P_R$ at each iteration is suitable for this purpose. Note that this cannot guarantee monotonic decreasing of the objective function in general, which may lead to non-optimal solutions. However, the simulation results shows that this approach is fairly robust and effective. The successful recovery of $P$ can be achieved as long as the initial guess for $P$ is not far from the optimal one. Since algorithm~\ref{algo1} can provide an estimate quite close to the true value, the requirement for the initial guess for $P$ thus would be met by using the fixed-point-based algorithm as an initialization.

Now we turn to develop the update rules for $P_L$, $P_R$ and $C^{gg,R}_\ell$ by alternately updating each matrix with keeping the other matrices fixed. For the details of the derivation of these update rules we refer to Appendix. Finally, the NMF-based algorithm dedicated to self-calibrate the photo-$z$ scatters from realistic measurements is specified in Algorithm~\ref{algo2}. 

\begin{algorithm}[h]
  \caption{The NMF-based algorithm to estimate $P$  from noisy data by solving for the problem of Eq.~\ref{eq:cggmin}. This algorithm should be run over many times with different starting points so as to find the globally optimal estimate for $P$.  Note that the appendix elaborates on the notations and conventions.} 
  \label{algo2} 
\begin{algorithmic} 
\STATE \COMMENT {\hrulefill\ Solving for Eq.~\ref{eq:cggmin}\hrulefill\.}
\REQUIRE{Non-negative data matrices $C^{gg,P}_\ell$ for all $\ell$.} 
\STATE {\bf initialize}: \\
(a) Assign $P$ from Algorithm~\ref{algo1}\\
(b) Compute $C^{gg,R}_\ell$ with Eq.~\ref{eq:c}   
\REPEAT 
 
\STATE 1. Update $W$: fixing $H_\ell$, updating $W$ with Eq.~\ref{eq:w}
\STATE 2. Update $H_\ell$: imposing $H_{\ell} =C^{gg,R}_\ell W^T$   
\STATE 3. Update $C^{gg,R}_\ell$: fixing $W$, updating $C^{gg,R}_\ell$ with Eq.~\ref{eq:c}
\UNTIL {a convergence criterion is satisfied (e.g. the change of each element in $W$ is below $10^{-8}$ or the maximum number of iterations is reached, $n_{\rm iter} =3\times10^{5}$)}


\ENSURE{$P$ (as $P\equiv W^T$) and $C^{gg,R}_\ell$} 

\end{algorithmic}
\end{algorithm}
 
We have to point out that, the optimization problem of Eq.~\ref{eq:cggmin} is not convex in $P$ and $C^{gg,R}_\ell$ together, which means that our proposed algorithm can only guarantee, if at all, to converge to a local minimum. Hence the estimated parameters are somewhat initial condition dependent, and good initializations can significantly eliminate unwanted solutions that can lead to large decomposition errors. Thanks to Algorithm~\ref{algo1} which would achieve a near global minimization successfully, in practice, we try to run this NMF-based optimization solver over many times with different initializations, and use the best minimum found (i.e., the minimum $\cal J$) as the global minimum to determine the globally optimal estimate of $P$.

\section{tests with mock data}\label{sect:test}

\begin{figure*}[htpb]
\centering
\mbox{
 \subfigure[photo-$z$ scatters]{
   \includegraphics[width=3.1in] {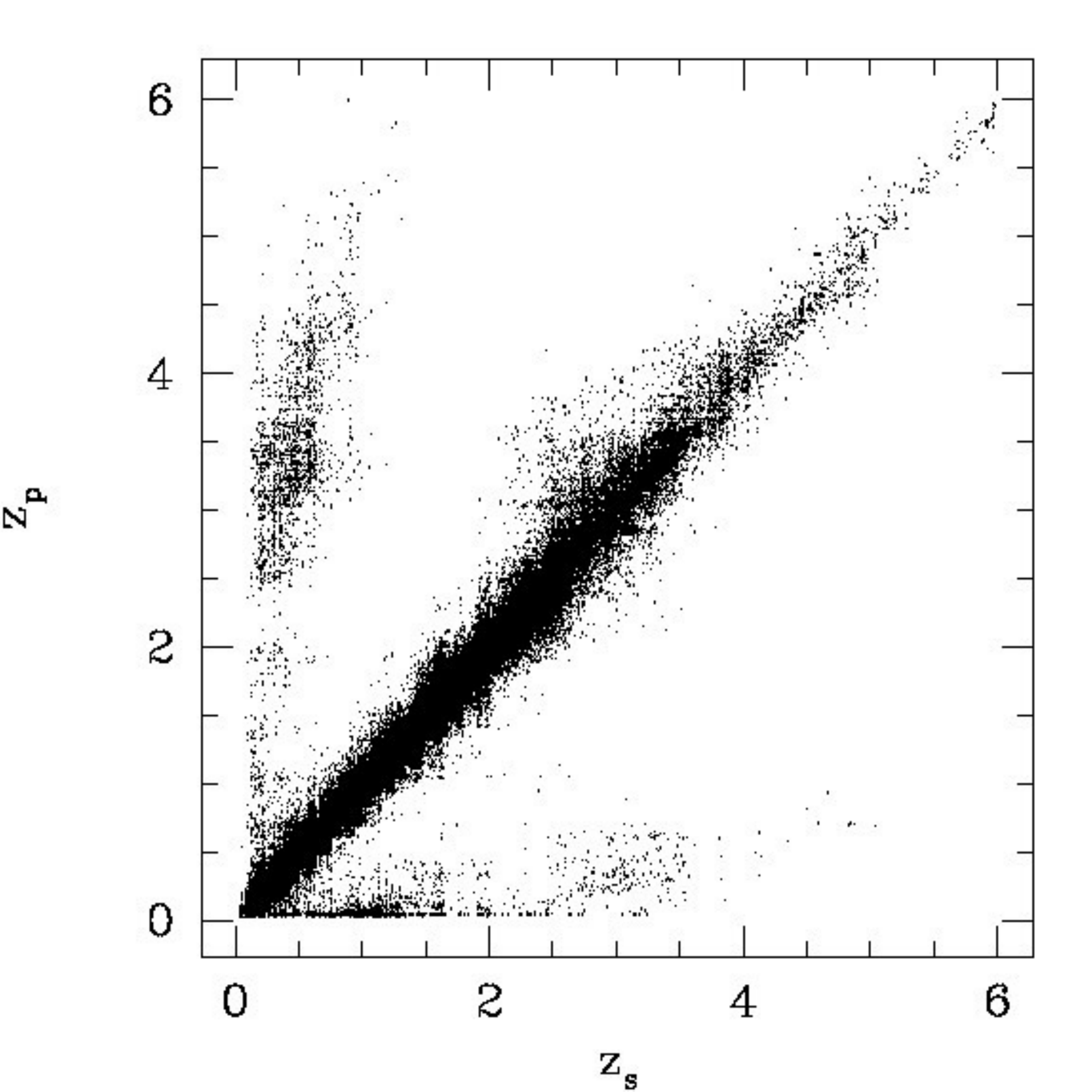}
    }

\subfigure[binned photo-$z$ scatters]{
   \includegraphics[width=3.1in] {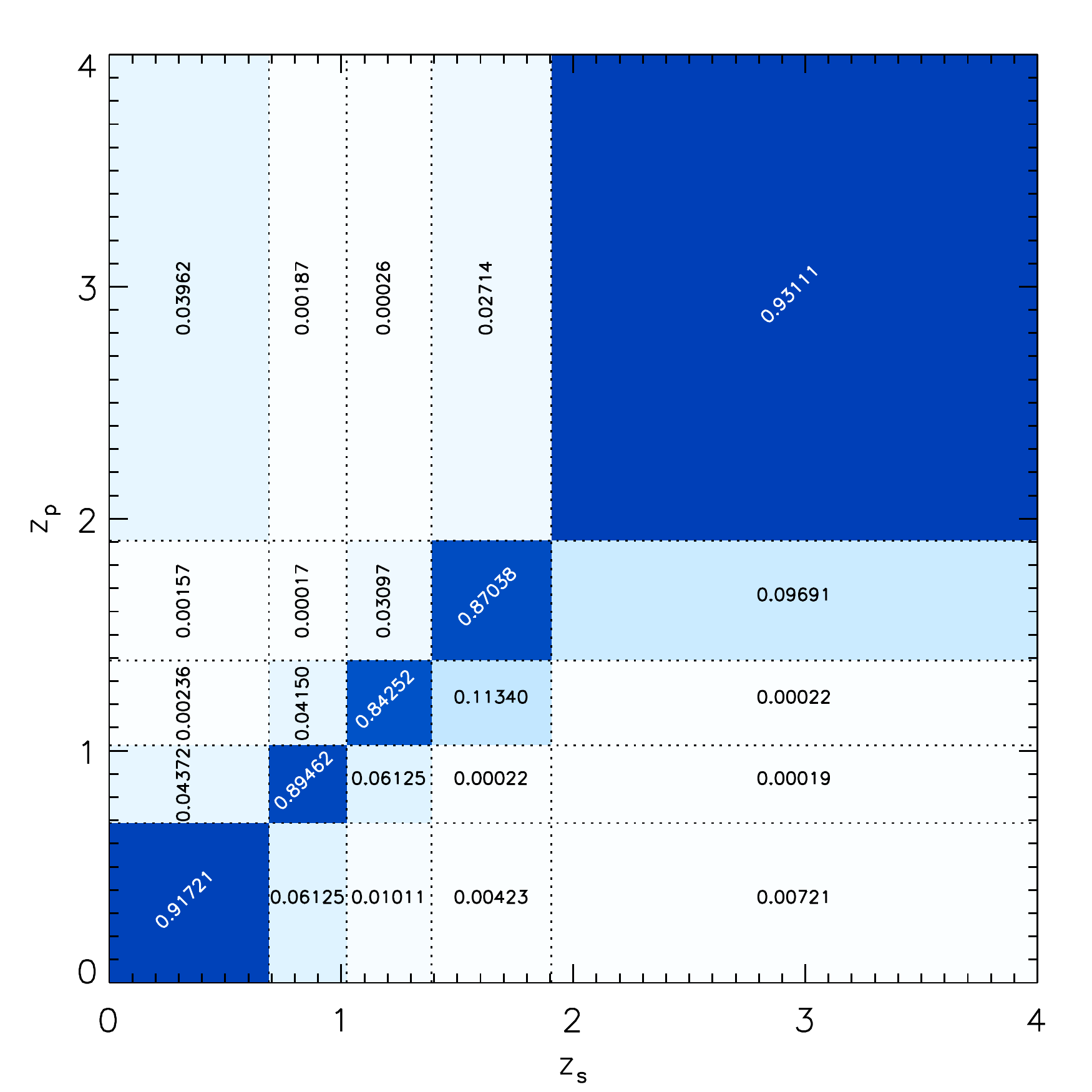}
   }
}
\caption{Photometric vs. spectroscopic redshift for the fiducial mock photometric set \citep{2010MNRAS.401.1399B}, where $z_S$ denotes the spectroscopic redshift, approximately corresponding to the true redshift, and $z_P$ denotes the photometric redshift. {\bf Left panel}: the fiducial photo-$z$ scatters in the simulated data~\citep{2010MNRAS.401.1399B} with 177210 galaxies. {\bf Right panel}: the corresponding bin-averaged photo-$z$ scatters $P_{ij}$, with 5 $z$-bins ranging from $0\leq z<4$. The values of all elements in the matrix $P$ are shown in each of the cells accordingly.}
\label{fig:p}
\end{figure*}

In this section, we test our proposed algorithms through mock data mimicking stage IV weak lensing experiments in order to demonstrate their effectiveness and robustness in self-calibrating photo-$z$ scatter in weak lensing surveys. There exist several exciting galaxy survey proposals in ``Stage IV'' lensing surveys such as Euclid, LSST and  WFIRST.  We mostly follow the data set and experimental settings in~\cite{2010MNRAS.405..359Z} (and references therein), whose survey area, galaxy number density and  redshift distribution are similar to that of LSST.

Following~\cite{2006MNRAS.366..101H} and~\cite{2006astro.ph.11159Z}, the fiducial galaxy redshift distribution $n(z_P)$ is chosen to have the form 
\beq\label{eq:nzp}
n(z_P) = x^2\exp(-x)dx/2\,,
\eeq 
where $x\equiv z_P/z_0$ with $z_0=0.45$. 

As mentioned in Sect.~\ref{sect:opt}, we have to appropriately choose the bin-widths of $\Delta \ell$ and $\Delta z$ such that an almost identical shot noise level both at each $\ell$-bin and each $z$-bin in the measured galaxy-galaxy power spectrum. Although the number of $z$-bins can be chosen arbitrarily, ~\cite{2006ApJ...636...21M} shows that we cannot improve the statistical errors on cosmological parameter constraints with $n \geq 5$. Hence we choose the following binning scheme throughout the paper. Using Eq.~\ref{eq:nzp}, we divide the galaxies into 5 $z$-bins at the range of $0\leq z < 4$ as $\{[0,0.688)$, $[0.688,1.023)$, $[1.023,1.389)$, $[1.389,1.906)$, $[1.906,4.0)\}$, each bin containing an equal number of galaxies, while we divide the total $\ell$-range $20\leq\ell<1000$ into $N_\ell=6$ spectra bands: \{$[20,408)$, $[409,577)$, $[578,707)$, $[708,816)$, $[817,913)$, $[814,1000]$\}, approximately with a constant $\ell \Delta \ell \simeq 82729$ for each bin.

To determine the noise covariance matrix $\sigma^{gg,P}_{ij}$ and $\sigma^{Gg,P}_{ij}$ with Eq.~\ref{eq:nzp}, we adopt a fiducial value $\bar{n}_g = 40$ per arcmin$^2$ and $f_{\rm sky} =0.5$. The rms dispersion in the shear measurement induced by the galaxy intrinsic ellipticities is adopted as $\gamma_{rms} =0.2$. For each simulation, the realizations of shot noise fluctuations are generated by using such noise model.

The fiducial scatters are obtained based on the simulated data from \cite{2010MNRAS.401.1399B}, which is produced for a SNAP-like survey using the method described in~\cite{2009A&A...504..359J}. Fig.~\ref{fig:p} illustrates the original photo-$z$ scatters in this data and the corresponding binned scatters adopted for this study. The left panel shows the $z_P$-$z_S$ distribution produced by the data having 177210 galaxies, where $z_S$ denotes the spectroscopic redshift. As we can see, the photometric redshifts in some island-shaped regions are grossly misestimated, leading to catastrophic redshift errors. For example, an ``island'' is at $z_P > 2.5$, $z_S < 0.6$, which is probably due to confusion between high-$z$ Lyman breaks and low-$z$ 400-nm breaks.

According to above binning scheme, the corresponding binned scatters $P_{ij}$ are shown at the right panel in Fig.~\ref{fig:p}. We have discarded the galaxy samples with $z_P>4$ in the data since we know such galaxies have large photo-$z$ catastrophic errors, while such galaxies merely account for $<1\%$ of the total and neglecting them should not lead to loss of information. The purpose of this study is to reconstruct all $P_{ij}$ using our proposed algorithms.

To generate the fiducial power spectra $C_{\ell}^{gg,R}$ and $C_{\ell}^{Gg,R}$, we use the public code CLASS~\citep{2011arXiv1104.2932L} and adopt a flat $\Lambda$CDM cosmology with $\Omega_m=0.268$, $\Omega_{\Lambda}=1-\Omega_m$, $\Omega_b=0.045$, $\sigma_8=0.85$, $n_s=1$ and $h=0.71$. For simplicity, we assume the fiducial galaxy bias to be $b_g = 1$.

\begin{figure}[t]
\centering
\includegraphics[width=3.1in] {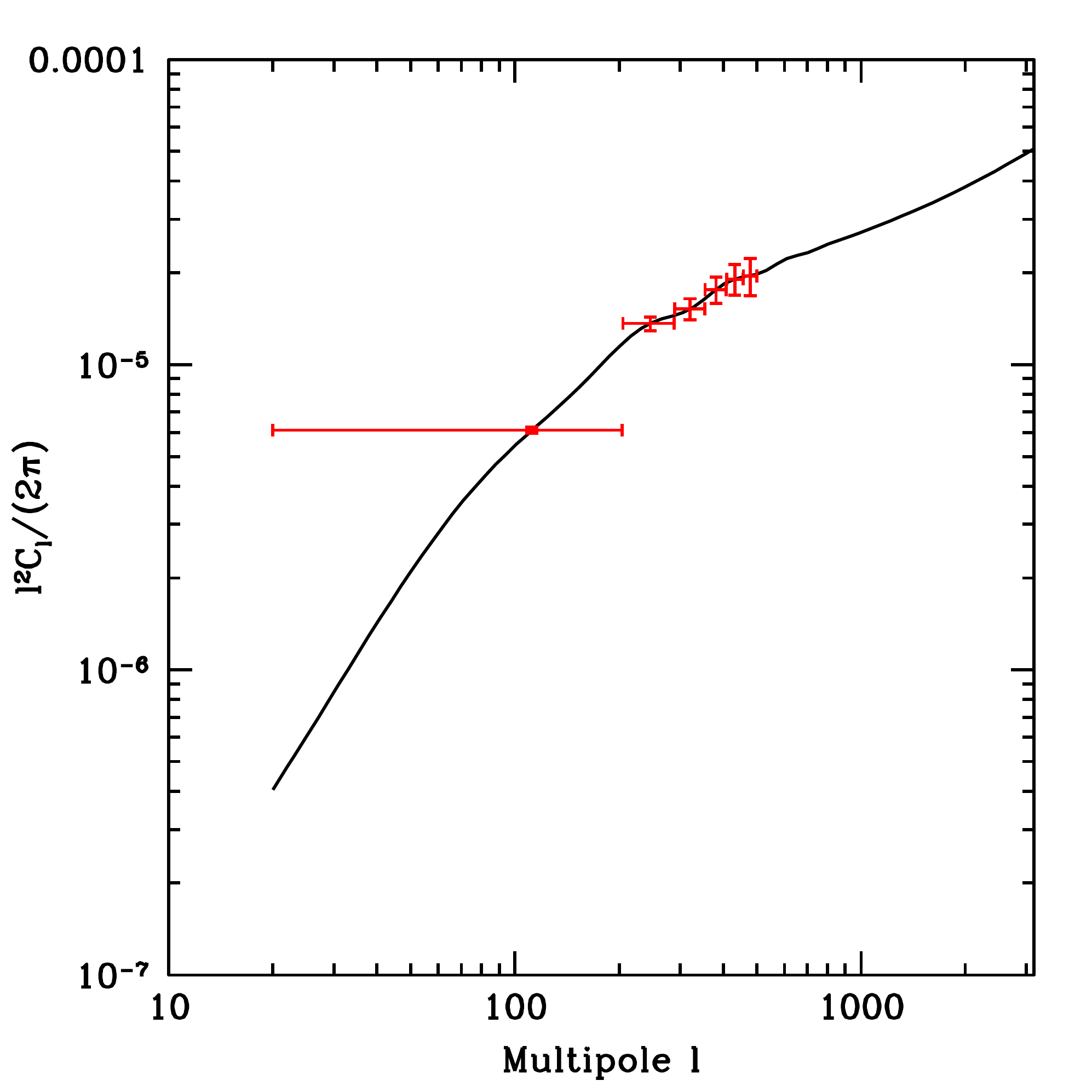}
\caption{Shot noise vs. non-zero cross-correlation $C_{i\neq j}^{gg,P}$ induced by photo-$z$ scatters. We show an example of the power spectrum $C_{35}^{gg,P}$ (black) between the photo-$z$ samples at the redshift bins of $[1.023,1.389)$ and $[1.906,4.0)$. Using the fiducial binning scheme, we show the resulting band-powers (red) in 6 $\ell$-bins and the associated 1-$\sigma$ error bars for comparison. Those statistical uncertainties arising from shot noise fluctuations are computed using Eq.~\ref{eq:noise1}.}
\label{fig:shot}
\end{figure}

Under the above simulation settings, an example of the resulting cross-power spectrum and its statistical uncertainty arising from the shot noise fluctuations is shown in Fig.~\ref{fig:shot}. As expected, the photo-$z$ errors induce a non-zero, but quite small cross-correlations between galaxies at different redshift bins. The associate uncertainties across all $\ell$-bins are almost identical as desired, with $\big(\sigma^{gg,P}_{i\neq j}\big)^2 = 3.68\times 10^{-11}$. However, the resulting signal-to-noise ratio, which is [91.9, 27.8, 18.9, 14.2, 11.7, 9.9] in each bin, tends to be low since the power spectrum of the signal increases with multipole $\ell$.   

\subsection{Simulation results}
Here we present the simulation results which were performed on a PC with 2.8GHz CPU and 16 GB RAM in python.

\subsubsection{Noise-free case}

\begin{figure}[h]
\centering
\includegraphics[width=3.1in] {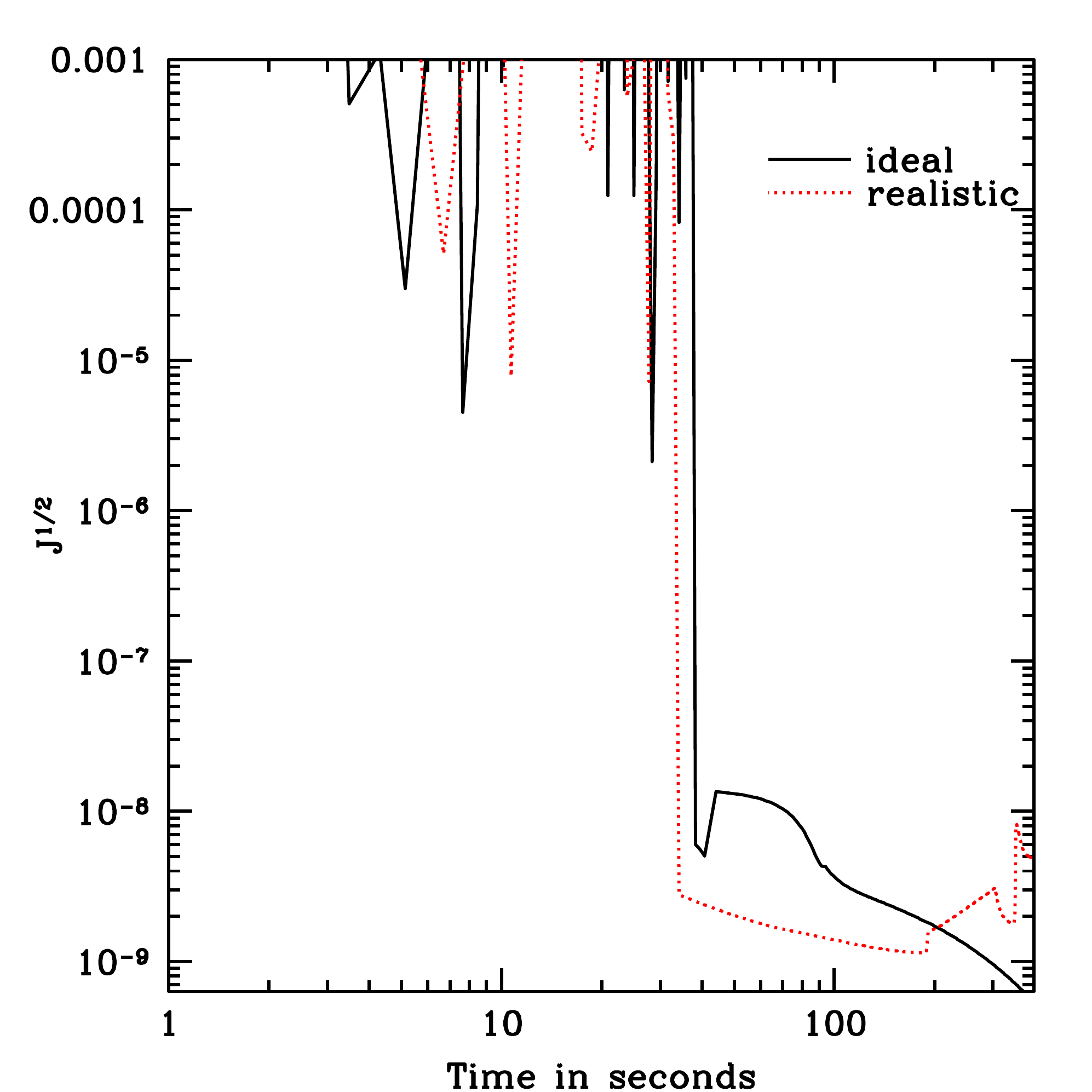}
\caption{Time vs. objective value ${\cal J}$ in Eq.~\ref{eq:cggmin} for the fixed-point-based algorithm applying to the fiducial data with (red) and without (black) shot noise. In each case, we show a typical run with a random initialization for scattering matrix $P$. Note that, in the realistic case, the objective value suddenly starts to increase, when ${\mathcal{J}}^{1/2}\simeq 10^{-9}$ which leads the recovery of each element of $P$ at the level of at least $\sim10^{-2}$. This is because this algorithm cannot guarantee the convergence for noisy data, but can derive a near-optimal estimate for $P$. In practice, we will use the NMF-based algorithm instead when this situation happens, and this near-optimal estimate is used for its initialization.}
\label{fig:time}
\end{figure}

We first evaluate the algorithm~\ref{algo1} on the idealized noise-free data, which is based on fixed-point iterations. To illustrate its fast convergence, Fig.~\ref{fig:time} checks the relation between the running time and the objective value. As can be seen, there are significant oscillations at the beginning stage of iteration, since we randomly generate a positive dense matrix $P$ as an initialization. Shortly after that, the updates become very efficient, yielding a sharply decrease on $\mathcal{J}$, and then the algorithm quickly decreases the objective function almost monotonically. Finally, the objective function converges to $\mathcal{J}^{1/2} \leq 5.8\times 10^{-10}$ by taking about 400 seconds and 300000 iterations, which makes an accurate recovery of each element of $P$ at the level of $\sim10^{-4}$ compared with its true value. Fig.~\ref{fig:time} therefore clearly demonstrates that, in the absence of shot noise, the fixed-point-based algorithm can efficiently solve the self-calibration problem exactly. Furthermore, it also illustrate that this algorithm can be successfully used in noisy data. Although this algorithm is designed for noise-free data, it can still  provide a near-global optimum solution with $\mathcal{J}^{1/2} \simeq 10^{-9}$, which may explain why this algorithm is applied for initialization of Algorithm~\ref{algo2} when estimating parameters from realistic measurements. 

\subsubsection{Realistic case}
\begin{figure*}[ht]
\centering
\mbox{
\subfigure[average value: $\left<P\right>$]{
   \includegraphics[width=3.1in] {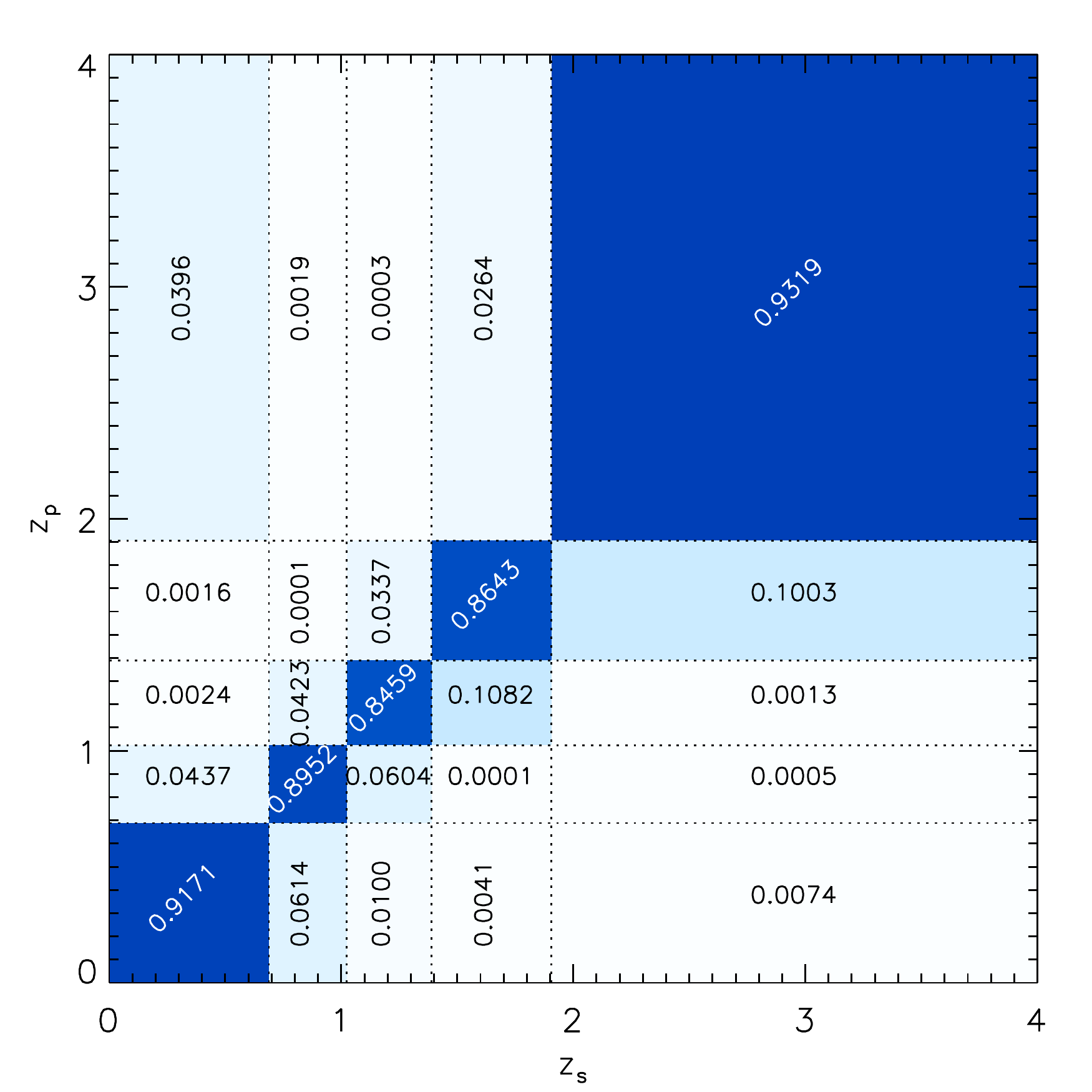}
  }
}  
\mbox{

\subfigure[absolute bias: $\left<|P-P^{\rm true}|\right>$ ]{
   \includegraphics[width=3.1in] {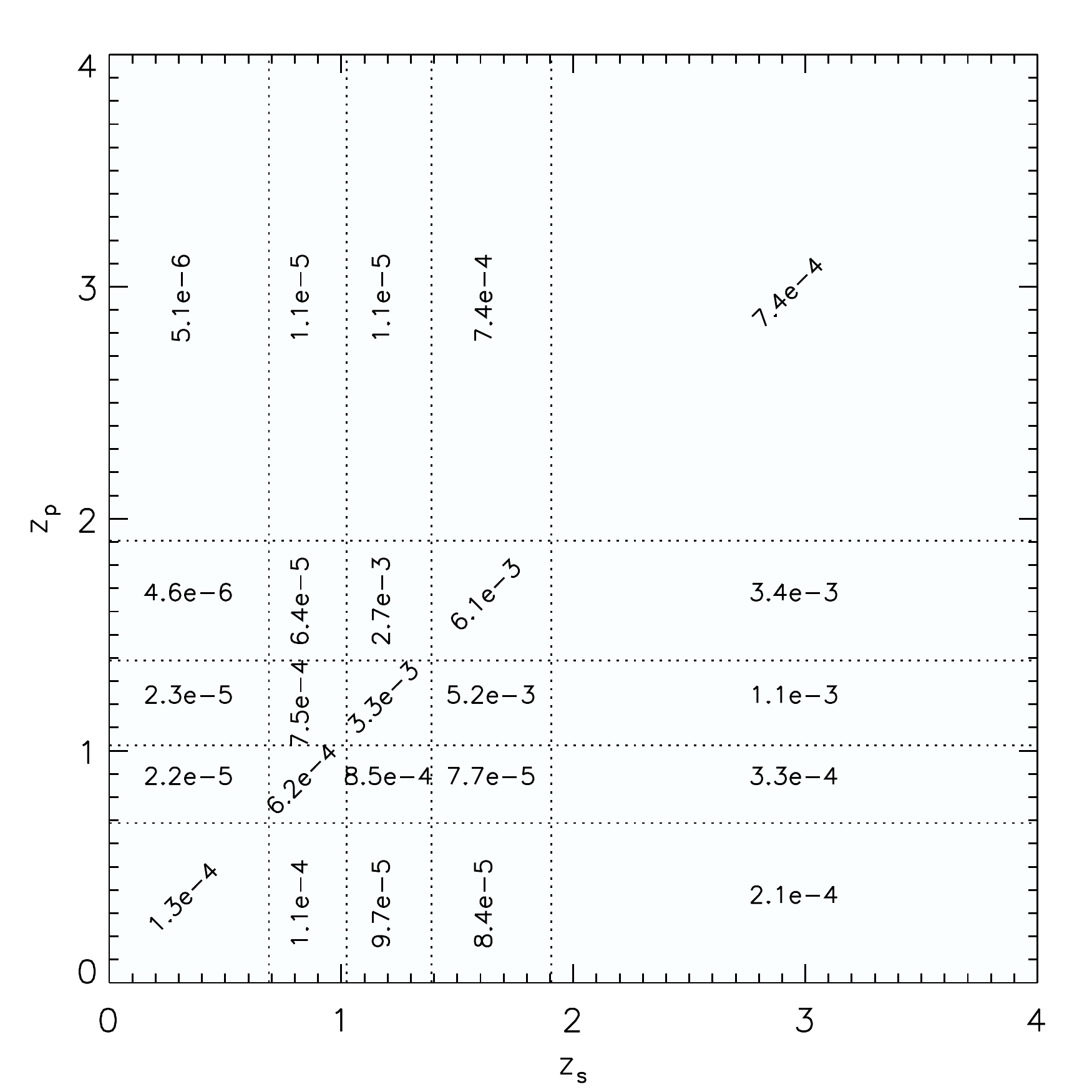}
    }

\subfigure[standard deviation: $\sigma_{P}$]{
   \includegraphics[width=3.1in] {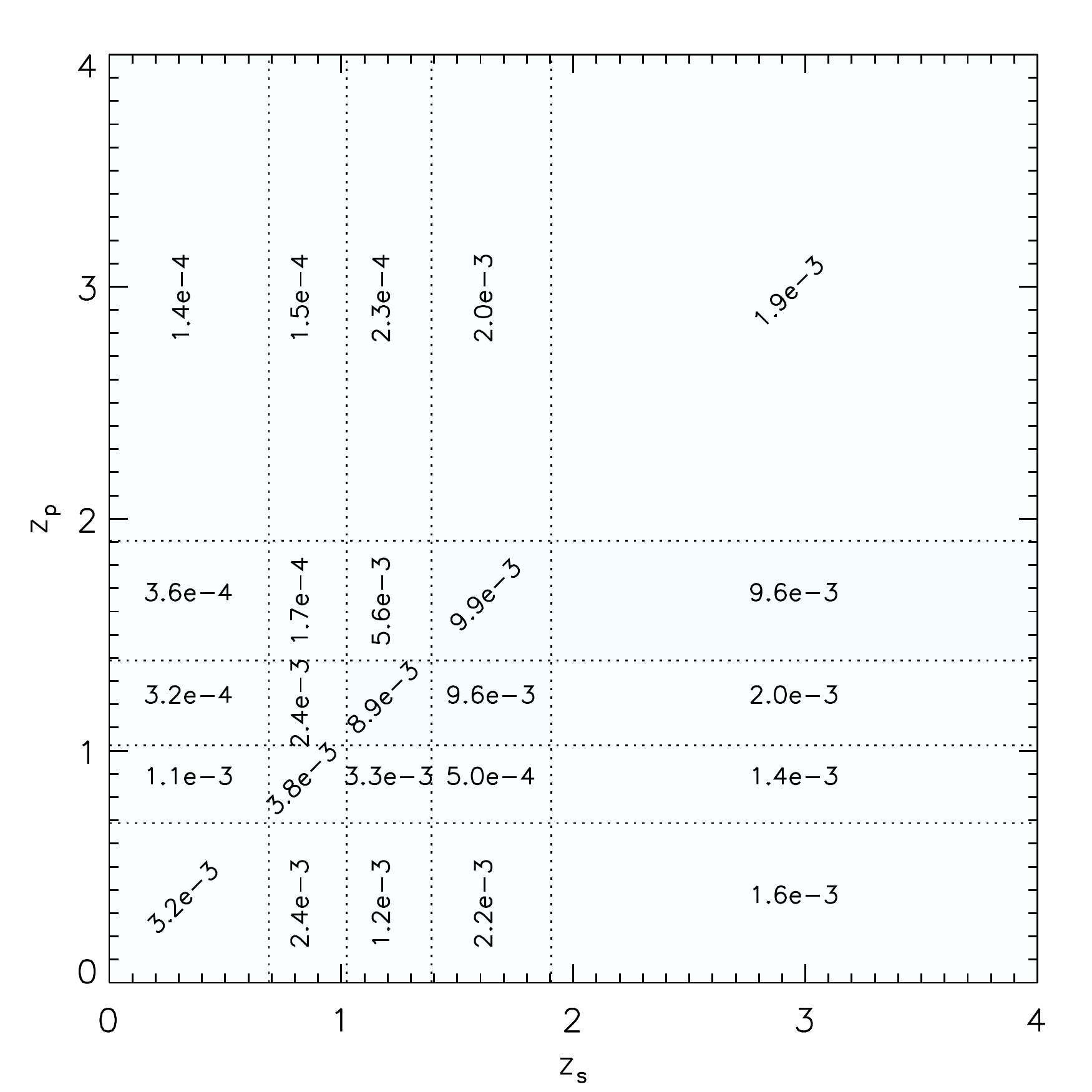}
   }
   }
 
\caption{Average value, absolute bias and standard deviation of the reconstructed scattering matrix $P$ obtained by averaging over the estimates from 50 realizations of noisy data based on algorithm~\ref{algo2}. To show the bias and its standard deviation more clearly, the same information is also provided in Fig.~\ref{fig:pmatrix}.} 
\label{fig:Pinfo}
\end{figure*}

\begin{figure}[htpb]
\centering
\includegraphics[width=3.2in] {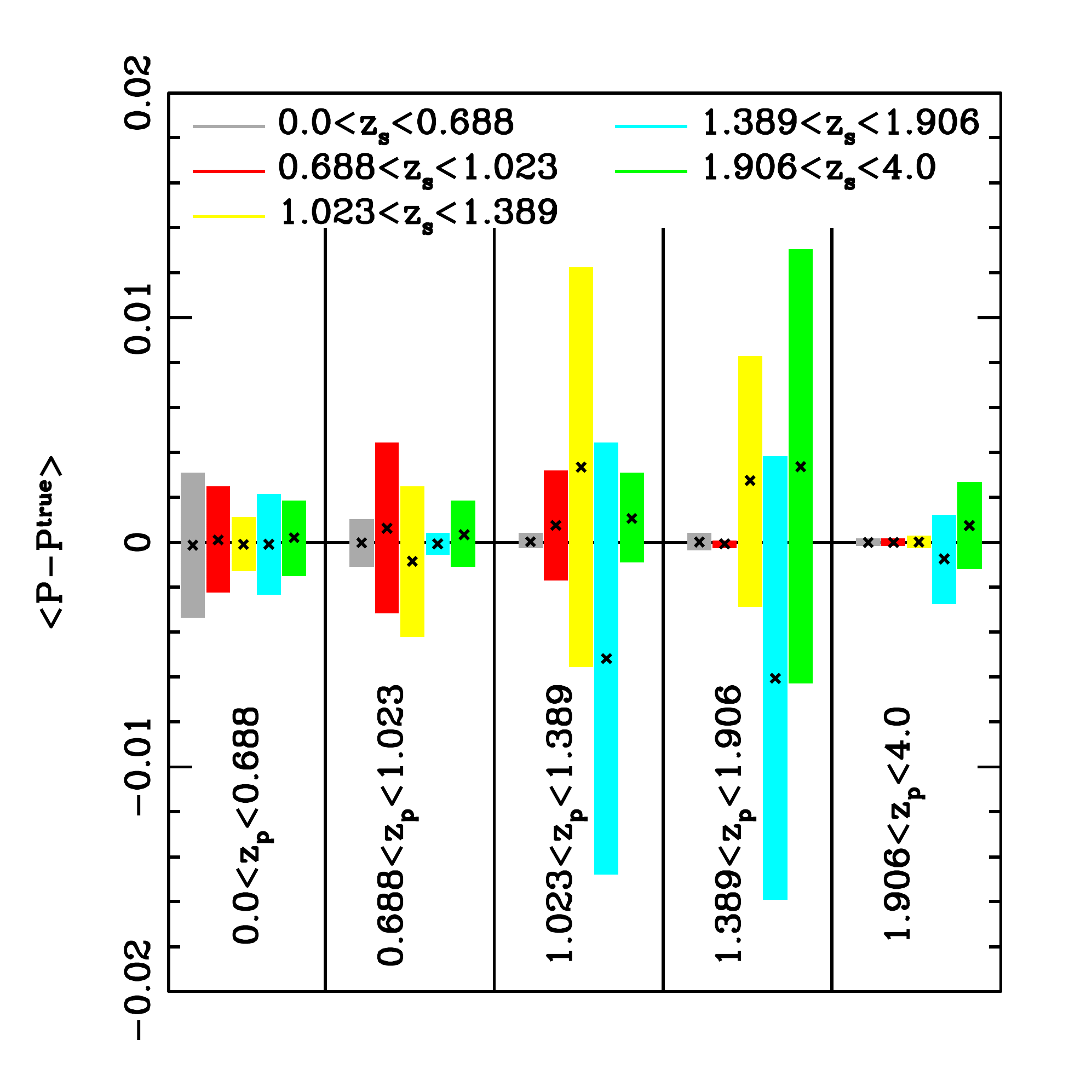}
\caption{Same as Fig.~\ref{fig:Pinfo}, but with a different view for the mean bias (represented as cross symbol) and its standard deviation of each element in the reconstructed scattering matrix $P$. } 
\label{fig:pmatrix}
\end{figure}

We now report performance and efficiency of algorithm~\ref{algo2} for noisy data. In order to investigate the reconstruction accuracy for the scattering matrix $P$, we have applied this algorithm to 50 realizations of simulated data, each with an independent random noise realization and identical simulated signal power spectra. For each realization, we ran this NMF-based algorithm 500 times with random initializations, and picked the result with the smallest ${\cal J}$ as its best estimate.

\begin{figure}[h!]
\centering
\includegraphics[width=3.2in] {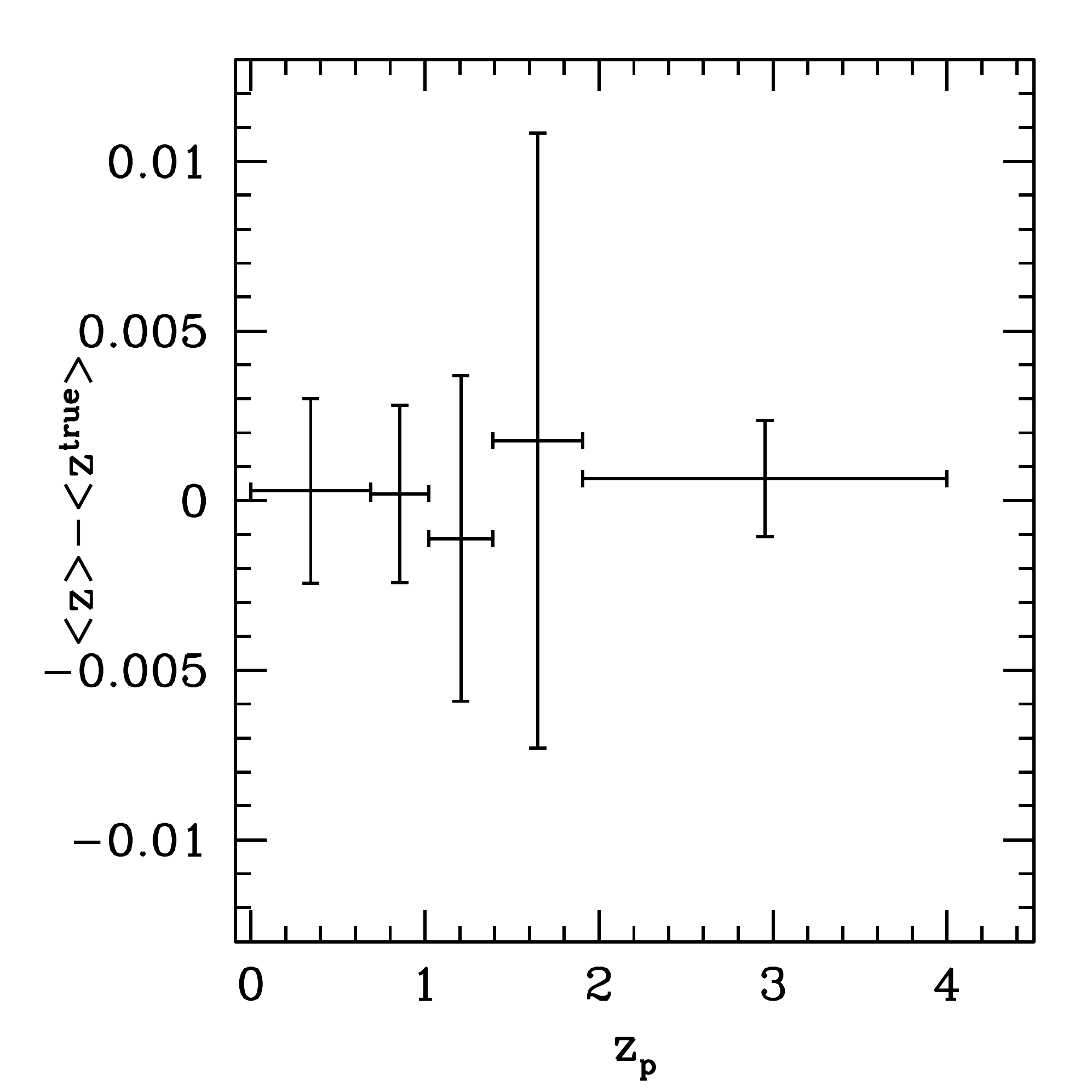}
\caption{Bias and associated standard deviation in the true mean redshift $\left<z\right>$ for the fiducial SNAP-like lensing survey. Both bias and standard deviation are determined from the reconstructed photo-$z$ scatters through Eq.~\ref{eq:zmean}, based on the 50 realizations of the simulated data.}
\label{fig:zmean}
\end{figure}

The ensemble-average scattering matrix, $\left<P\right>$, is obtained by averaging over such best estimates of the scatters from those realizations, and the associated standard deviation $\sigma_{P}$ is obtained from its dispersion. The corresponding absolute bias, $\left<|P-P^{\rm true}|\right>$, can be computed easily by comparing with the simulation truth. We aggregate all information about the scatters in Fig.~\ref{fig:Pinfo} to produce the average value, the absolute bias, and standard deviation. 

We observe that, the differences between the reconstructed matrix elements $P_{ij}$ and their respective true values are significantly small, about half of the elements reaching an accuracy level of $<1\times 10^{-4}$. Since the true values of the diagonal elements are close to unity, much larger than off-diagonals, the diagonals have some relatively large biases. Even for the worst case where the bias reaches the maximum value of $0.006$, the reconstruction accuracy is still much less than $1\%$. More interestingly, the scatters for some bins with $P^{\rm true} <10^{-3}$ can be reconstructed almost exactly, with extremely small reconstruction errors of $<10^{-5}- 10^{-6}$. Furthermore, the mean of all the derived biases (25 elements in total) is about $0.001$, which is sufficiently small and indicates that the proposed NMF-based algorithm almost provides an unbiased estimate for scatters in the average sense.

In addition, the statistical errors of the reconstructed scatters, $\sigma_P$, are also quite small, reaching the accuracy of 0.003 on average. Since the shot noise fluctuations at diagonals are larger than off-diagonals by a factor 2, as expected by Eq.~\ref{eq:noise1} in our fiducial model, the uncertainties of diagonal elements are somewhat larger than those of off-diagonals. 

Finally, we conclude that, the NMF-based algorithm is able to successfully detect the scattering probabilities with high accuracy at the level of $10^{-3}$, which may meet the accuracy requirement for the fiducial ``stage IV'' lensing survey.

Moreover, we can use a single convenient number, the density weighted true redshift for each photo-$z$ bin, to quantify the quality of the reconstruction for photo-$z$ errors (see~\cite{2010MNRAS.405..359Z} for details). It can be well approximated by    
\beq\label{eq:zmean}
\left<z_i\right> \simeq \sum_{j} P_{ji}\left<z_j^P\right> =\sum_{j} P_{ji} \frac{\int_j z_P n(z_P)dz_P}{\int_j n(z_P)dz_P}  \,,
\eeq
where $\left<z_i\right>$ is the density weighted true redshift of the $i$-th photo-$z$ bin, $\left<z_j^P\right>$ stands for the density weighted average photo-$z$ of the $j$-th photo-$z$ bin, and $n(z_P)$ is defined in Eq.~\ref{eq:nzp}. Based on the reconstructed scatters $P_{ij}$ for the simulation data, the corresponding averaged bias and the statistical error in $\left<z_i\right>$ can be then computed accordingly. The results are summarized in Table 1 and illustrated in Fig.~\ref{fig:zmean}. We can see that, the bias value in each photo-$z$ bin is negligible small, ranging from $2\times10^{-4}$ to $0.002$, which is of order $0.03\% - 0.3\%$ of the true redshift. Meanwhile, the associated statistical error appears nearly $3-15$ times greater than the bias value in each bin, but still well under percent accuracy. Both biases and statistical errors slowly vary with the redshifts $z_P$.

Note that all the simulation results are based on the fiducial data at low-$\ell$ range, $20\leq\ell<1000$, whereas collecting more information on high multipoles would improve the determination of $P$. If high-$\ell$ measurements of the power spectra are taken into account, there will be more parameters from high-$\ell$ bins to be estimated, while high-$\ell$ data will tend to give low signal-to-noise ratio. When solving the non-convex optimization problem numerically, those two factors might decrease the reconstruction accuracy in practice and increase the computational time. More detailed tests for the inclusion of high-$\ell$ data will be made in future work.

\begin{table}[h]
\begin{center}
\begin{tabular}{c|c|c|c|c}
\hline
$z_P$ range&  $\left<z_P\right>$ & $\left<z^{\rm true}\right>$ & $\left<z\right>-\left<z^{\rm true}\right>$ & $\sigma_{\left<z\right>}$  \\
\hline
$[0.0 - 0.688]$ & 0.472& 0.521  &$3\times 10^{-4}$& 0.003    \\
$[0.688 -1.023]$&0.857  &  0.861  & $2\times 10^{-4}$&  0.003   \\
$[1.023 - 1.389]$&1.199 &  1.232 & $-$0.001 &  0.005 \\
$[1.389 - 1.906]$&1.625 &  1.693  & 0.002 &  0.009   \\
$[1.906 - 4.0]$&2.485 & 2.378  &$6\times 10^{-4}$ & 0.002  \\ 
\hline
\end{tabular}
\caption{mean bias and associated standard deviation of the reconstructed density weighted true redshift, estimated from 50 simulations varying noise realizations.}
\end{center}
\label{tab:z}
\end{table}

\section{Conclusions}\label{sect:con}
Based on the fixed-point and the non-negative matrix factorization methods, we have developed an iteration algorithm to solve the constrained nonlinear optimization problem arising from the studies of the self-calibration of photo-$z$ scatters in weak lensing surveys. The algorithm was applied to the mock data mimicking a  stage IV project, including galaxy density-density correlations and the density-shear cross-correlations which are contaminated by shot noise.

Our proposed algorithms exploits the ``fixed-point'' iteration and ``multiplicative update rules'' to efficiently minimize the objective function, leading to find globally optimal estimates. The algorithm has the great virtue of being remarkably stable, robust and fast. The typical run for finding the optimal solution takes about 15 minutes for the noisy data. 

The results are very promising. For the noise level in a stage IV survey like LSST, photo-$z$ outlier rates can be determined at an accuracy level of $10^{-3}$ on average, even if we only use the information at $\ell<1000$.  This leads to nearly unbiased estimates of the true mean redshift of each photo-$z$ bin, at the level of $10^{-3}$ for the bins over the redshift range from 0 to 4. Such precision would almost attain the desirable statistical accuracy of the future ``stage IV'' projects, allowing for precision weak lensing cosmology.

The high quality of the reconstruction indicates that the algorithm proposed in this paper is pretty suitable for weak lensing cosmology where one wants to accurately determine and understand the catastrophic photo-$z$ error rates, and for understanding the photometric redshift. The algorithm so far only applies to data at $\ell<1000$. When necessary, we should also include the information at $\ell>1000$ to further improve photo-$z$ self-calibration. Since it is computationally challenging, we will leave the investigation elsewhere.

Of course, still much work has to be performed to better examine the robustness and effectiveness of those algorithms. It will be tested under more realistic simulated observations, with taking into account some possible extra error sources that might slightly degrade the reconstruction accuracy. These include the magnification bias and the size bias in the galaxy number distribution,  unphysical correlation induced by observational selection effect, as well as the intrinsic cross correlation between adjacent galaxy bins. For example, the magnification bias can also cause spatial correlation between foreground and background galaxies, which may be misinterpreted as photo-$z$ errors.  This problem of magnification bias can be dealt with by utilizing the unique flux dependence of magnification bias on galaxy flux~\citep{2010MNRAS.405..359Z}. For a given photo-$z$ bin, we further split galaxies into several flux bins. The magnification bias in these flux bins is $\propto g(F)$ which is an observable determined by the observed galaxy flux distribution function. Since this $g(F)$ dependence is in general different to the dependence of galaxy bias on $F$, we can separate the magnification bias from the galaxy intrinsic clustering~\citep{2017arXiv170301575Y}.  A simpler but less optimal alternative is to weigh galaxies by their flux appropriately. As long as we require $\sum_i w_ig_i=0$, the weighted overdensity $\delta_g^W\equiv \sum_i w_i \delta_g^i$ is free of magnification bias. Here $g_i$, $\delta^i_g$ and $w_i$ are the $g$ factor, galaxy number overdensity, and the weighting factor of the $i$-th flux bin. The algorithm developed in the current paper then directly applies to $\delta_g^W$. Nevertheless, the existence of magnification bias (and size bias) complicates the photo-$z$ self-calibration. It is an important issue for further investigation. Also, it is necessary to test it against and apply it onto real data in future work.

\section*{Acknowledgments}
This work was supported by the National Science Foundation of China (11433001, 11320101002, 11621303, 11403071), National Basic Research Program of China (2015CB85701), a grant from Science and Technology Commission of Shanghai Municipality (Grants No. 16DZ2260200), Key Laboratory for Particle Physics, Astrophysics and Cosmology, Ministry of Education, and Shanghai Key Laboratory for Particle Physics and Cosmology (SKLPPC). We would like to thank the anonymous referee for valuable suggestions which helped us to significantly improve this paper.

\bibliography{le}
\bibliographystyle{apj}
\nocite{*}

\appendix
\section{}
\vspace{-1.5em}
We here present a derivation of the expressions for the update rules of $P_L$, $P_R$ and $C_\ell^{gg,R}$ in Eq.~\ref{eq:minimize}. Those rules will be constructed by alternately updating each matrix with keeping the other matrices fixed.
\subsection{Derivation of the update rules for $P_L$, $P_R$}

In the following analysis, we will use the alternative notations $V_\ell\equiv C_\ell^{gg,P}$, $W\equiv P_L^T$ and $H_{\ell}\equiv C^{gg,R}_\ell P_R$ to clearly describe the algorithm without loss of generality. We seek to the update rules for decreasing the objective function 
\beq
{\cal J} = \frac{1}{2}\sum_\ell \norm{V_\ell - W H_{\ell} }_F^2\,.
\eeq

Based on the gradient-descent technique similar to that used in~\cite{Lee01algorithmsfor}, in what follows, we show the update rule for the $W$ with keeping $H_\ell$ fixed. By Taylor expanding the objective function ${\cal J}$ with respect to the element $w_{ab}$ in $W$ about the point $w^t_{ab}$, we obtain 
\beq
{\cal J}(w) = {\cal J}(w^t_{ab}) + {\cal J}'_{w_{ab}}(w-w^t_{ab}) + \frac{1}{2}{\cal J}''_{w_{ab}}(w-w^t_{ab})^2\,,
\eeq
where ${\cal J}'_{w_{ab}}$ and ${\cal J}''_{w_{ab}}$ are the first and second derivative with respect to  $w_{ab}$, respectively, and the high order derivatives are zeros. Thus it is easy to check that 
\beqs
\frac{\partial{\cal J}}{\partial w_{ab}} &=& -\sum_l \left[(V_\ell-WH_\ell)H^T_\ell\right]_{ab}\,,\\
\frac{\partial^2{\cal J}}{\partial^2 w_{ab}} &=& \sum_l \left[H_\ell H^T_\ell\right]_{bb}\,,
\eeqs

which obviously for positive elements has the upper bound

\beq
\sum_\ell [H_\ell H^T_\ell]_{bb}\leq \sum_\ell \frac{[WH_\ell H^T_\ell]_{ab}}{w_{ab}}\,.
\eeq

Then we can define an auxiliary function for the update rule. The auxiliary function regarding $w_{ab}$ is defined as 
\beqs
G(w,w_{ab}) = {\cal J}(w_{ab}) + {\cal J}'_{w_{ab}}(w_{ab})(w-w_{ab}) \\
+ \frac{1}{2}\sum_\ell \frac{[WH_\ell H^T_\ell]_{ab}}{w_{ab}} (w-w_{ab})^2\nonumber\,.
\eeqs

To make ${\cal J}$ to be nonincreasing under the update, we construct  
\beq
{\cal J}(w^{t+1}_{ab})\leq G(w^{t+1}_{ab},w^t_{ab})\leq G(w^t_{ab},w^t_{ab}) = {\cal J}(w^t_{ab})\,,
\eeq
where the first inequality comes from the upper bound and the second is the result of minimization. Setting  
\beq
\frac{\partial{G(w,w_{ab})}}{\partial w}=0\,,
\eeq

we finally obtain the update rule for $W$:
\beq
w_{ab}=w_{ab} \frac{\sum_\ell [V_\ell H_\ell^T]_{ab}}{\sum_\ell[WH_\ell H^T_\ell]_{ab}}\,,
\eeq
which is consistent with the ``multiplicative update rules'' for the standard NMF, when $n_\ell=1$.

Next, this update rule can be easily extended to accommodate the constraint of the row-wise unitary sum for $W$ (recall $W\equiv P_L^T$ and the column-wise unitary sum for $P_L$). Introducing a set of Lagrangian multipliers, $\lambda_i$ with $i=1,\ldots,n$, we can rewrite the objective function as 
\beq
\tilde{\cal J} = {\cal J} + \sum_i\lambda_i(1-\sum_j W_{ij})
\eeq 
Similar to the proof by \cite{DBLP:conf/scia/ZhuYO13}, we find that the objective function is nonincreasing by using the following multiplicative update rule
\beq\label{eq:w}
W_{ij} \leftarrow W_{ij} \frac{\nabla^-_{ij} A_{ij}+1-B_{ij}}{\nabla^+_{ij}A_{ij}}\,,
\eeq 
with $\nabla^-_{ij}= \sum_\ell [V_\ell H_\ell^T]_{ij}$, $\nabla^+_{ij}=\sum_\ell[WH_\ell H^T_\ell]_{ij}$,  $A_{ij} = \sum_b W_{ib}/\nabla^+_{ib}$, $B_{ij} = \sum_bW_{ib}\nabla^-_{ib}/\nabla^+_{ib}$. Note that there is a negative term $-B_{ij}$ in the numerator, which may cause negative entries in $W$ during updates. Hence we apply the ``moving term'' trick~\citep{DBLP:journals/tnn/YangO10} to overcome this problem if any negative entries in $W$ appear, which gives $W_{ij} \leftarrow W_{ij} \frac{\nabla^-_{ij} A_{ij}+1}{\nabla^+_{ij}A_{ij}+B_{ij}}$.

Once the update for $W$ (i.e. $P_L^T$) is obtained, we update $P_R$ by imposing $P_R = W^T$ such that the constraint $P_L = P_R$ is implemented during iterations.

The simulation tests shows that the use of the following convergence criterion is appropriate: the change of each element in $W$ per iteration satisfies $|W^{t}_{ij} -W^{t+1}_{ij}| <10^{-8}$ or the maximum number of iterations $n_{\rm iter} =3 \times 10^{5}$ is reached.

\subsection{Derivation of the update rules for $C^{gg,R}_\ell$}

Let us now consider the update rule for $C^{gg,R}_\ell$ with $W$ fixed. Based on the fact that $C^{gg,R}_\ell$ for all $\ell$ are diagonal matrices, Eq.~\ref{eq:cggmin} thus can be rewritten in matrix-vector form as
\beq
{\cal J } = \sum_\ell \norm{V_\ell -W  C^{gg,R}_\ell W^T}_F^2 = \sum_\ell \norm{ \textrm{Vec}\left[V_\ell\right] -U \vec{c}_\ell}_F^2\,,
\eeq
where the ``vectorization'' operator ($\textrm{vec}[\cdot]$) is used for converting the matrix into a column vector by stacking the columns into a long column vector, and $\vec{c}_\ell$ collects all of the diagonal elements of $C^{gg,R}_\ell$, i.e., $[\vec{c}_\ell]_i =[C^{gg,R}_\ell]_{ii}$ for all $i$. Here $U$ is the $n^2 \times n$ matrix, consisting of the Kronecker products, denoted by $\otimes$, on the columns of $W$ as its columns, i.e., $U = [\vec{w}_1 \otimes \vec{w}_1,\ldots,\vec{w}_n \otimes \vec{w}_n]$, where $\vec{w}_i$ is the $i$-th column of $W$. Obviously, according to a well-known solution for this linear least squares problem, the optimal $\vec{c}_\ell$ can be found as
\beq\label{eq:c}
\vec{c}_\ell = (U^TU)^{-1}U^T \textrm{Vec}\left[V_\ell\right],\quad \textrm{for all}~\ell\,,
\eeq
which determines the optimal $C^{gg,R}_\ell$ and, in other words, is of course equivalent to its update rule. To ensure the non-negativity, an additional step is added after each such update to project all negative elements of $C^{gg,R}_\ell$ to be their absolute values. Note that, it is not necessary to use NMF technique for updating $C^{gg,R}_\ell$, since the objective function is convex in $C^{gg,R}_\ell$ and this update rule is stable and efficient for decreasing objective function monotonically.


\end{document}